# Visualizing the low-energy electronic structure of the triplet superconductor UTe$_2$ through quasiparticle interference

Anuva Aishwarya[1,2]*, Hans Christiansen[3], Sheng Ran[4,5,6], Nicholas P. Butch[4,5], Brian M. Andersen[3], Andreas Kreisel[3], Shanta R. Saha[4], Johnpierre Paglione[4,7], Vidya Madhavan[1,7]*

[1]Department of Physics and Materials Research Laboratory, University of Illinois at Urbana- Champaign, Urbana, IL, USA

[2]Department of Physics, Harvard University, Cambridge, MA, USA

[3]Niels Bohr Institute, University of Copenhagen, DK-2100 Copenhagen, Denmark

[4]Maryland Quantum Materials Center, Department of Physics, University of Maryland, College Park, MD, USA

[5]NIST Center for Neutron Research, National Institute of Standards and Technology, Gaithersburg, MD, USA.

[6]Department of Physics, Washington University in St. Louis, St. Louis, MO 63130, USA

[7]Canadian Institute for Advanced Research, Toronto, ON M5G 1Z8, Canada

*Correspondence to: <u>aaishwarya@fas.harvard.edu</u>, <u>vm1@illinois.edu</u>

**Abstract**
The identification, control and theoretical modelling of spin-triplet superconductors (STC) remain a central theme in quantum materials research. Intrinsic STC are rare but offer rich condensate properties and unique surface properties allowing insights into the nature of the spin-triplet order, and promising applications in quantum technologies. Owing to interactions, the order parameter in STCs can often be intertwined with other symmetry breaking orders like charge/spin density waves (CDW/SDW) or pair density waves (PDW) complicating their phase diagrams. UTe$_2$ stands out as the only known odd-parity, STC that harbors such intertwined orders on the surface and possible topological surface states composed of Majorana fermions. While the (0-11) facet is the most heavily studied, the fermiology of this surface that gives rise to such exotic phenomena is still lacking and continues to be an area of active interest. Here, we employ low-temperature spectroscopic imaging to reveal the Fermi surface of UTe$_2$ through quasiparticle interference. We find scattering originating from the uranium-derived bands that play a major role in the formation of the CDW and the PDW phases. Tunneling spectroscopy further reveals spectral signatures of the CDW gap, corroborating its onset temperature. Suppressing the CDW with a magnetic field, highlights the presence of small, circular Fermi pockets that disperse strongly near the Fermi energy. We discuss the nature of the interference patterns and the origin of the small Fermi pockets in the context of the calculated band structure and the unconventional CDW phase.

There has been long-standing search for spin-triplet superconductors which harbor fundamentally new excitations like Majorana fermions and half-quantum vortices. Cooper pairs in a spin-triplet superconductor possess angular momentum S=1 and are more robust to external magnetic fields. Since the spin part of the Cooper pair wavefunction is symmetric under exchange, the remaining quantum numbers must necessarily be antisymmetric. Thus, the superconducting order parameter is odd parity in momentum space, i.e. p-wave or f-wave and unconventional, as opposed to the well understood s-wave state that forms the basis of Bardeen-Cooper Schrieffer (BCS) theory. An odd-parity condensate has been established in superfluid 3He [1], but it is not a superconducting state of paired charged particles. Unambiguous evidence for odd-parity superconductivity in material platforms was lacking (except possibly UPt$_3$ [2,3]), until the discovery of superconductivity in UTe$_2$ in 2019 [4], accompanied by strong evidence for spin-triplet pairing that has stood the test of time.

The superconducting state in UTe$_2$ is exotic, with the highest Pauli limit at 65 T [4,5] known for any low- T$_c$ superconductor. The superconducting phase diagram in a magnetic field is rich and highly anisotropic, with reentrant and field-polarized superconductivity [5] existing in different parts of the phase diagram. The Knight shift measurements reveal only a 5% reduction across Tc along the a-direction and minimal suppression (less than 0.2%) along b and c axes, indicating a spin-triplet order parameter (OP) [4,6] with an anisotropic d-vector. There is no consensus on the exact momentum structure of the superconducting order parameter (OP) with putative candidates being a nodal [7-9] or chiral OP [10,11]. To add to this stew of unconventional phenomena, UTe$_2$ hosts symmetry-breaking intertwined orders within the superconducting state, including a magnetic-field sensitive CDW order [12] and a PDW order, possibly arising from strong electronic correlations [13]. The CDW breaks time reversal and mirror symmetries under an applied magnetic field and is suppressed near the H$_{c2}$ of the SC order [14]. This leads to the hypothesis that the CDW is caused by a parent PDW, which is a superconducting state that breaks translational symmetry in addition to the global U(1) symmetry [15]. Since SC and CDW are both instabilities of the Fermi surface, and seem to be strongly connected in UTe$_2$, understanding the unique ground state of UTe$_2$ requires a thorough knowledge of the momentum space structure of its Fermi surface. Importantly, the necessary and sufficient conditions of a Fermi surface that give rise to such a spin-triplet superconductor is an intriguing fundamental question that remains to be answered.

Angle resolved photoemission spectroscopy (ARPES) experiments have indicated that the key players involved in the low-energy fermiology are the uranium d- and f- bands and the tellurium-derived p- bands [16]. We show such bands from a tight-binding model in Fig.1f together with the rectangularly warped cylindrical Fermi surfaces in momentum space (Fig.1e). Quantum oscillation measurements agree with the presence of cylindrical Fermi surfaces centered around the X and Y points derived from the U-d bands and Te-p bands [17, 18]. The effective masses extracted from these measurements are of the order 30-80 m$_e$ indicating strongly renormalized, massive bands possibly arising from Kondo interactions [17-19]. Although different quantum oscillation measurements agree with the cylindrical nature of Fermi surfaces, they have reached contradictory conclusions on the existence of small 3D Fermi pockets derived from f-bands. Owing to the three-dimensional nature of the crystal, difficulty in getting micron-sized cleave surfaces and the presence of correlated f-electrons, obtaining a clear picture through ARPES has proved difficult. Also unanswered is an interesting open question pertaining to the role of the f-electrons and the Kondo correlations in the superconducting state of UTe$_2$. Scanning tunneling microscopy and spectroscopic (STM/S) imaging (also known as quasiparticle interference (QPI) imaging) is a suitable probe to address these questions, as it can work with sub-micron sized cleaved terraces while delivering excellent energy and momentum-space resolution. In this work, we employ high-resolution spectroscopy and spectroscopic imaging at T = 300 mK (below Tc of UTe$_2$) to visualize the low temperature Fermi surface in UTe$_2$ that gives rise to the exotic superconducting state and all its ancillary orders.

UTe$_2$ is an orthorhombic crystal that crystallizes in the *Immm* space group [20]. There are two crystallographically inequivalent Te sites (labelled Te1 and Te2 in Fig.1a) based on the relative bond length of U-Te. Cleaving single crystals at 90 K reveals flat terraces of the (0-11) facet as shown by the large area topography in Fig. 1b. The surface is made of rows of Te1 and Te2 atoms along the crystallographic a direction as shown by the high-resolution atomistic topography in the inset of Fig. 1b. Uranium atoms are sub-surface and lie between the Te1 and Te2 atoms, as shown in the schematic. The lattice of UTe$_2$ is a centered rectangular lattice (shown by the schematic in Fig. 1b) and falls into the *cm* wallpaper group characterized by horizontal reflections that go through lattice points and horizontal glide reflections which lie midway between lattice points. The primitive lattice vectors are shown in the inset of Fig. 1b by the arrows. This lattice gives rise to 6 Bragg peaks in the Fourier transform (FT) and a hexagonal shaped Brillouin zone (BZ) as shown in Fig. 1c. The six lattice Bragg peaks within this formalism lie at $(q_x, q_{c^*}) = (0, \pm\frac{2\pi}{c^*})$, $(\pm\frac{2\pi}{a}, \pm\frac{\pi}{c^*})$ where $a$ is the Te-Te separation along the chain and $c^*$ is the distance between the Te-chains. For a detailed description of this lattice in real space and reciprocal space, we refer to Ref. [27]. Note that $q_x$ aligns with the $k_x$ direction in the momentum space for UTe$_2$ and $q_{c^*}$ is the direction on the surface perpendicular to $q_x$. Using the above vectors we construct the surface BZ (SBZ) marked by the dashed hexagon, which captures all the momentum space points of the (0-11) projection of the bulk orthorhombic BZ of UTe$_2$ [27]. A slice of the local density of states (LDOS) at -10 meV is shown in Fig. 1c. We can identify several peaks in the FT, apart from the lattice Bragg peaks, which reside close to the SBZ. Peaks labelled 1,2 and 3 correspond to $q_{1,2}^{CDW} = (\pm 0.43\frac{2\pi}{a}, \frac{\pi}{c^*})$, $q_3^{CDW} = (0.57\frac{2\pi}{a}, 0)$ which have been previously associated with a CDW state [12,13].

In this work, we study UTe$_2$ single crystals doped with 1% thorium. Thorium is routinely used as a substitutional defect in U-based compounds as it is closest in atomic size to uranium and is not expected to introduce additional strain or disorder. The lack of QPI-related signal in the parent-compound is not unusual, as d band states are known to be challenging to image. This is primarily due to matrix-element effects arising from the limited overlap of the primarily d-wave and f-wave Wannier orbitals of the sample with the s-wave orbital of the tip [28]. Moreover, the number of scattering sites associated with the U-atoms is small in the parent compound. The thorium substitution provides additional scattering sites to the Bloch electrons leading to enhanced interference. We detect the presence of bright triangular sub-surface defects in the large area topographies (Fig. 1b), with a relative concentration of about 1.2% which agrees with the nominal concentration of thorium in the bulk single crystal. In high-resolution topographies where we can see both the Te sublattices, these defects reveal a clover leaf shape (shown in the atomistic topography in inset of Fig. 1b and supplement), which is the characteristic symmetry of the underlying uranium site. Uranium is sub-surface and shares bonds with neighboring Te1 and Te2 atoms. Given the location, we expect that these defects originate from either a uranium vacancy or a thorium substitution. Resistivity and magnetization measurements as a function of temperature resemble the parent compound, featuring no anomalies above 2 K. Bulk characterization through transport and susceptibility measurements findicate that at 1% thorium substitution, Tc is smaller (Tc ~ 0.95-1 K). [supplement, 21,22]. We confirm from dI/dV spectroscopy and spectroscopic imaging that the minimal thorium substitution does not alter the low-temperature properties. The evidence for this is two-fold. First, below about 30 K, Kondo screening of the f-orbital derived moments by the itinerant bulk bands renders UTe$_2$ into a strongly-correlated Kondo metal. This many-body metallic state is characterized by a resonance near E$_F$ which manifests as a Fano peak in dI/dV spectroscopy as shown in Fig 2 (and supplement for average dI/dV spectrum). We indeed observe the characteristic Fano lineshape with a peak at about ~-5 meV similar to the parent compound shown in the supplement. Secondly, UTe$_2$ also hosts a low-temperature CDW order which manifests as additional peaks close to the SBZ boundary in the FT as observed in the FT of LDOS slice at -10 meV shown in Fig. 1c.

High-resolution spectroscopy along the Te chains (Fig. 2a) reveal, an additional feature around ~ -1 meV (Fig. 2b) inside the Fano resonance. This peak oscillates at the same wavevector as the CDW as shown by the intensity plot in Fig. 2c. This suggests that the partial gap-like feature is the spectral fingerprint of the CDW. (A similar gap can be observed clearly in the ultra-low temperature dI/dV spectrum obtained at T = 300 mK,[25]) The gap-like feature at ~ 1 meV gives a $T_{CDW} = \frac{2\Delta}{3.5 k_B} = 6.6$ K (using the BCS gap equation for CDW [23]) which agrees well with the measured CDW onset temperature of T$_{CDW}$~ 5- 6 K.

To study the normal-state electronic structure, we consider STS maps and their respective FTs at energies higher than the amplitude of the SC gap on the surface (2Δ ≈ 500 $\mu eV$). The FT of such STS maps details scattering wavevectors $q$, which connect $k_i$ and $k_f$ states on the contours of constant energy surface in momentum space thus providing information on the band structure. Figure 3 shows LDOS maps at representative energies obtained on a 50 nm by 50 nm area. If we consider Fig. 2a-d, and their respective FTs in 2e-h, we find that these FTs are distinct from the FT of the parent compound (supplement, [29]). Apart from the Bragg peaks and CDW peaks, we observe a strong signal at the BZ boundary that straddles the CDW peaks. Segments of these q-vectors are replicated in higher order BZs (indicated by arrows).

In order to understand the phenomenology of the QPI, we use a tight-binding model consistent with experimental findings in ARPES and quantum oscillations which contains two p-orbital states of Te and two states of the U atoms. The explicit form of the tight-binding parameters and orbital hybridization giving rise to the tube-like Fermi surfaces is discussed in Ref. [27, 30]. We construct the Green's function of the bulk and the surface using an iterative method, taking into account the experimentally relevant (0-11) cleaving surface, and using that momenta along the surface are good quantum numbers [27]. The resulting spectral function is shown in Fig. 3i for the bulk and Fig. 3j for the surface states at energies beyond the superconducting order parameter, displaying clear imprints of the U-dominated bands (green intensity), while the Te bands are spread out at all momenta giving a small background signal (not visible). The joint density of states (JDOS) calculated from the spectral function at $E_F$ is shown in Fig. 3k (full dataset in supplement). Intensities from scattering processes between the "green bands" in Fig. 3i and 3j are clearly visible in the JDOS. Note the periodicity of the hexagonal shaped SBZ as expected from the elementary cell of the surface. The comparison of the experimental data allows us to identify structures originating from the bulk electronic structure and contrast the effects of a CDW, which is not part of the simulations. Comparing the JDOS from the tight-binding calculations to our data, the QPI seen in STM likely arises from the scattering of electrons between cylindrical Fermi surfaces of the bulk U d-bands.

Next, we turn our attention to the LDOS and QPI maps close to the Fermi energy. Our previous work has already shown that a perpendicular magnetic field around 10 T is sufficient to suppress the CDW significantly at $E_F$. Fig. 4 shows an interesting QPI pattern emerging in the FTs of LDOS maps in a perpendicular magnetic field, B = 10.5 T as the CDW is strongly suppressed on either side of the $E_F$ (compare Fig 3 and Fig 4, full dataset in the supplement). The LDOS maps in field has been obtained on the same field of view as the 0 T maps with the same tip. We can summarize our observations under two main categories. First, we observe cross(X)-like spectroscopic features at the edge of the SBZ (indicated by yellow arrows in Fig. 4b-c). Second, we observe the emergence of circular pockets in Fig. 4c-d. These pockets only emerge close to $E_F$. The cross-like feature that we see in a magnetic field, results from two distinct components, indicated by the blue and yellow arrows. The component marked by the blue arrow is present at both 0 T and 10.5 T, while the one marked by the yellow arrow appears only at 10.5 T. Interestingly, these new features closely resemble what would be expected if the original QPI were 'folded' or 'unfolded' back into the SBZ.

Zooming in and comparing the zero field FT to the 10.5 T FT at $E_F$, we notice that the circular pockets always existed but were obscured by the CDW peaks and become apparent only when the peaks are suppressed. To visualize the evolution of these features as a function of energy, we compare cuts along specific momentum directions (Fig. 5a) and track them as a function of energy. Fig. 5b shows a cut along $q_{c^*} = -\frac{\pi}{c^*}$ along the CDW peak, and Fig. 5c shows a cut along the $q_c^* = 0$ in magnetic field. For $q_{c^*} = -\frac{\pi}{c^*}$, we notice a strongly dispersing signal near the CDW peak, $q_x = 0.43\frac{2\pi}{a^*}$, denoted by the dashed white line in Fig. 5b. This is the signal associated with the emerging circular pockets that we observe in the FT-STS. Given the outward dispersion, the nature of the bands from which the signal arises are electron-like.

Finally, we address the question of the origin of the dispersing electron-like small, circular pockets at the $(q_x, q_{c^*}) \sim (\pm\frac{\pi}{a}, \pm\frac{\pi}{c^*})$ and $(q_x, q_{c^*}) \sim (\pm\frac{1.1\pi}{a^*}, 0)$. These states do not have a simple interpretation in terms of superconducting topological surface states due to their energy scale and the fact that they are pivoted off the TRIM (Time-reversal invariant momentum) points of the surface BZ. One possibility is that they arise from scattering off impurities that are located in the layer below the surface. To explore this possibility, we show in Fig. 5f QPI simulations where the elementary cell is extended to the second layer "A" where impurities are placed [31]. As opposed to the JDOS, the QPI signal depends on the nature and location of the scatterer [31]. Compared to the case where the impurities are located in the surface layer "B" (Fig. 5e), one sees that the different position of the impurities impose a phase shift of the scattered electronic state, and therefore yields a QPI intensity pattern with maxima at larger $q_x$ for $q_{c^*} = 0$ and smaller $q_x$ for $q_{c^*} = \frac{\pi}{c^*}$ (Fig. 5f). This QPI simulation recovers the features of the experimentally observed QPI signal including the X-like feature. One can also identify three circular-like regions in Fig. 5f reminiscent of the circles in Fig. 5g. However, given that this intensity originates from scattering across the d-bands, which presumably disperse very weakly within the measured energy range, these intensities in QPI are not expected to move much in q-space within a few meV of energy (supplement). This is in contrast to the strong dispersion with energy across a few meV, of the small circular pockets observed by our experiment. Since the theoretical simulations do not incorporate CDW order, this brings us to another interesting possibility, namely that the circular bands originate from scattering across small Fermi pockets. The CDW can result in folding of bands resulting in small Fermi pockets, similar to the Chern Fermi pockets that have been recently observed in the kagome superconductors. In the kagome metals these Fermi pockets are concentrated sources of Berry curvature and orbital magnetic moment. Pairing of electrons across these Fermi pockets can give rise to emergent topological and correlated phases including a loop-current pseudogap phase and chiral topological PDWs [32]. Whether a similar mechanism is at play on the surface of UTe$_2$ remains an important future question to be answered. Additionally, the connection between the small Fermi pockets detected by STM/S in this study and the purported existence of a small bulk 3D Fermi surface observed in ARPES and quantum oscillations remains to be deciphered [16,17].

In conclusion, we report the visualization of the Fermi surface of UTe$_2$ through QPI imaging. We observe primary scattering originating from uranium d bands at the SBZ edges. This agrees with our JDOS simulations from theoretical calculations, which exhibit peaks at the location of the CDW peaks at all energies, pointing towards a divergence in susceptibility at these momenta. Additionally, we observe circular pockets near the CDW peaks that disperse strongly in energy, unlike the d-bands. Our QPI simulations can reproduce parts of the experimentally observed QPI using scattering at uranium sites from different layers. The significant dispersion of the states forming the circular pockets, however, leads us to conclude that additional band-folding effects of the CDW order may be important to quantitatively capture the electronic and topological properties of the surface states in the fascinating heavy-fermion material of $UTe_2$.

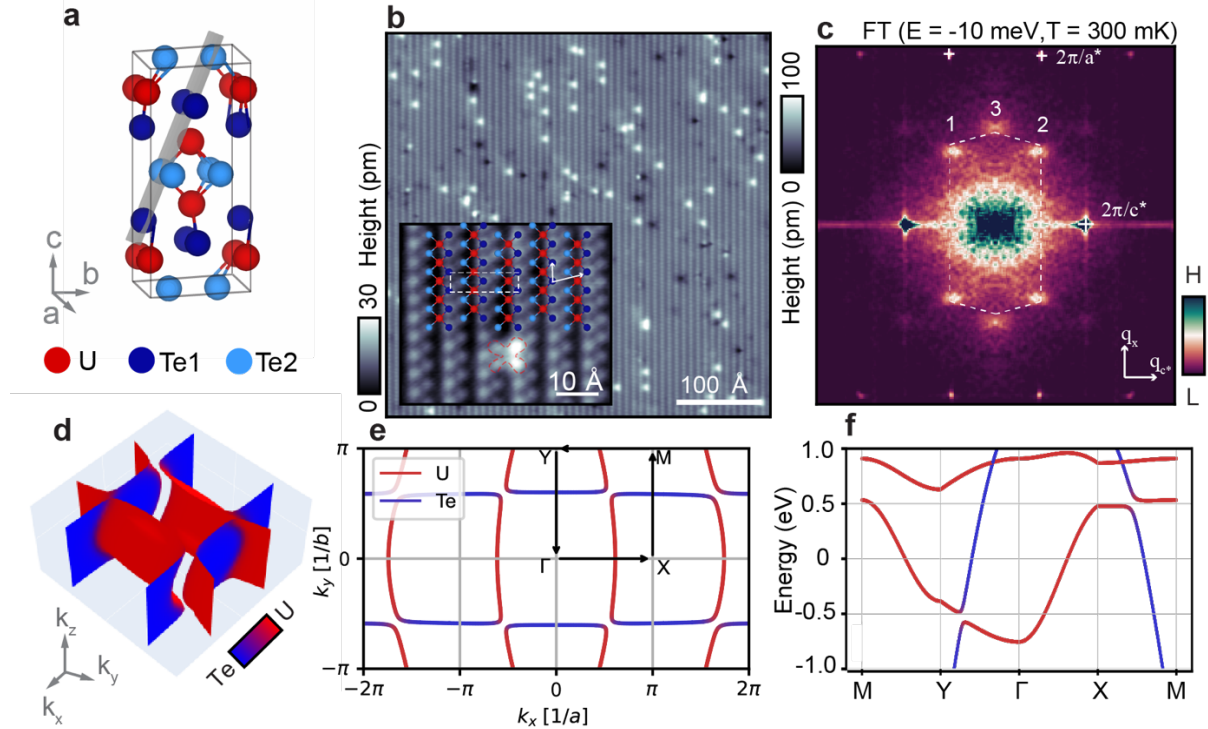

**Figure 1: Crystal structure, (0-11) surface, Brillouin zone and surface Brillouin zone a,** Unit cell of UTe$_2$ showing uranium atoms and tellurium atoms. The two different tellurium atoms are shown in different shades of blue. The grey shaded plane marks the (0-11) plane. **b** Topography of a typical field of view (45 nm by 45 nm) of the (0-11) cleave surface, with rows of Te atoms. The bright atoms are the sub-surface uranium defects or thorium impurities. (Inset, Atomically resolved topography with the Te1 and Te2 atoms overlayed with the schematic of the (0-11) surface. The schematic includes both the Te atoms and the U atom closest to the surface. The dashed rectangle shows the unit cell of the primitive lattice which is a centered rectangle and the white arrows are the primitive lattice vectors. Also included in the topography is a bright uranium site defect denoted by the dashed clover leaf shape. The defect brightens up 2 Te1 atoms and 2 Te2 atoms reflecting the symmetry of the uranium site.) **c,** Fourier transform of an LDOS map obtained at E =-10 meV, T = 300 mK (V$_{Bias}$= 50 mV, I = 250 pA), showing the surface Brillouin zone (indicated by the white dashed line). The crosses denote the reciprocal lattice vectors. The peaks labelled 1, 2 and 3 are the CDW peaks. **d,** Band structure of UTe$_2$ based on tight binding calculations that show cylindrical Fermi surfaces, where the colors indicate the orbital content of the bands. **e,** Slice of the Fermi surface at $k_z = 0$ depicting the high symmetry directions. The blue and red represent tellurium and uranium derived bands respectively. **f,** Dispersion of the bands across a large energy range along different high-symmetry directions of the BZ from the tight-binding calculations.

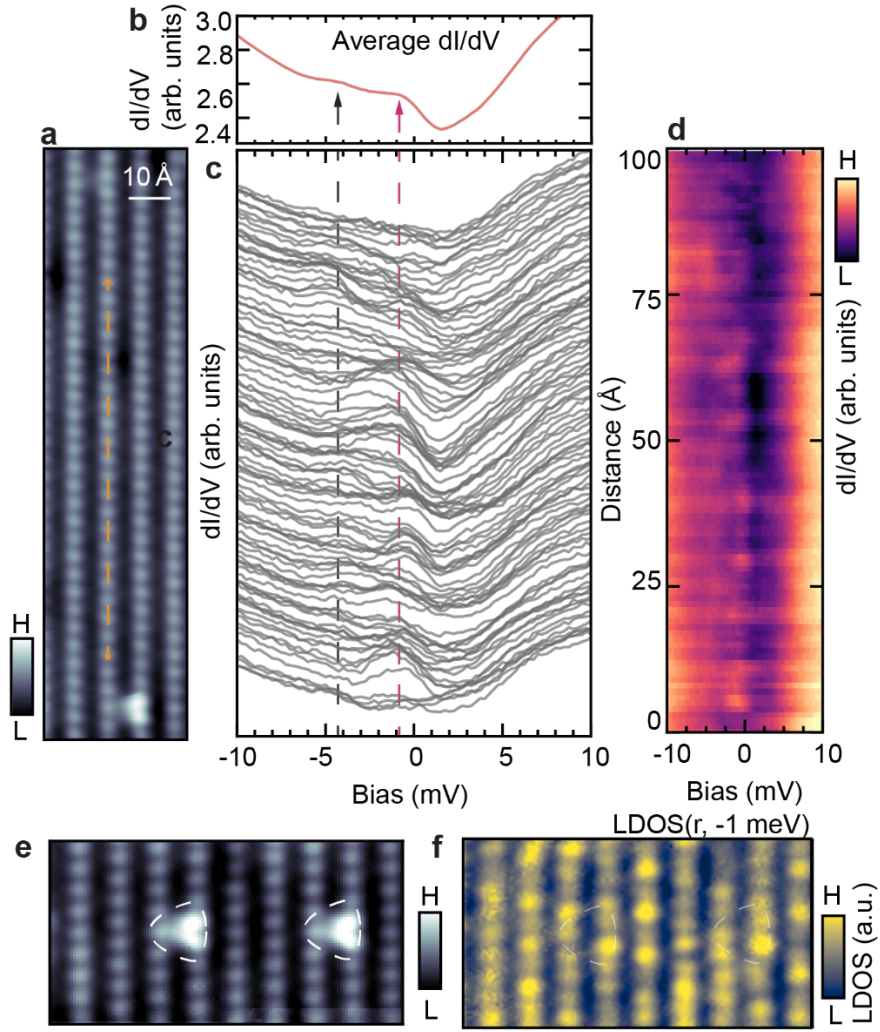

**Figure 2: Spectral fingerprint of the charge density wave and emergence of additional energy scale below the Kondo coherence**

**a,** Topography obtained at T = 300 mK showing the location of dI/dV linecut in orange dashed line. **b,** Average dI/dV linecut obtained over 100 spectra. **c,** Waterfall plot of the high resolution dI/dV linecut obtained along the Te chains. Black dashed line shows the Fano peak and orange dashed line shows the peak associated with the CDW gap. The dI/dV spectra have been offset for clarity to show the strong modulation of the peak at ($V_{Bias}$ = 50 mV, I = 300 pA, T = 300 mK). **d,** Intensity plot of the same linecut showing the clear modulation of the peak. **e,** Atomically resolved topography showing the characteristic U defects indicated by dashed white lines. **f,** Measured LDOS at -1 meV associated with the CDW gap ($V_{Bias}$ = 50 mV, I = 250 pA, T = 300 mK).

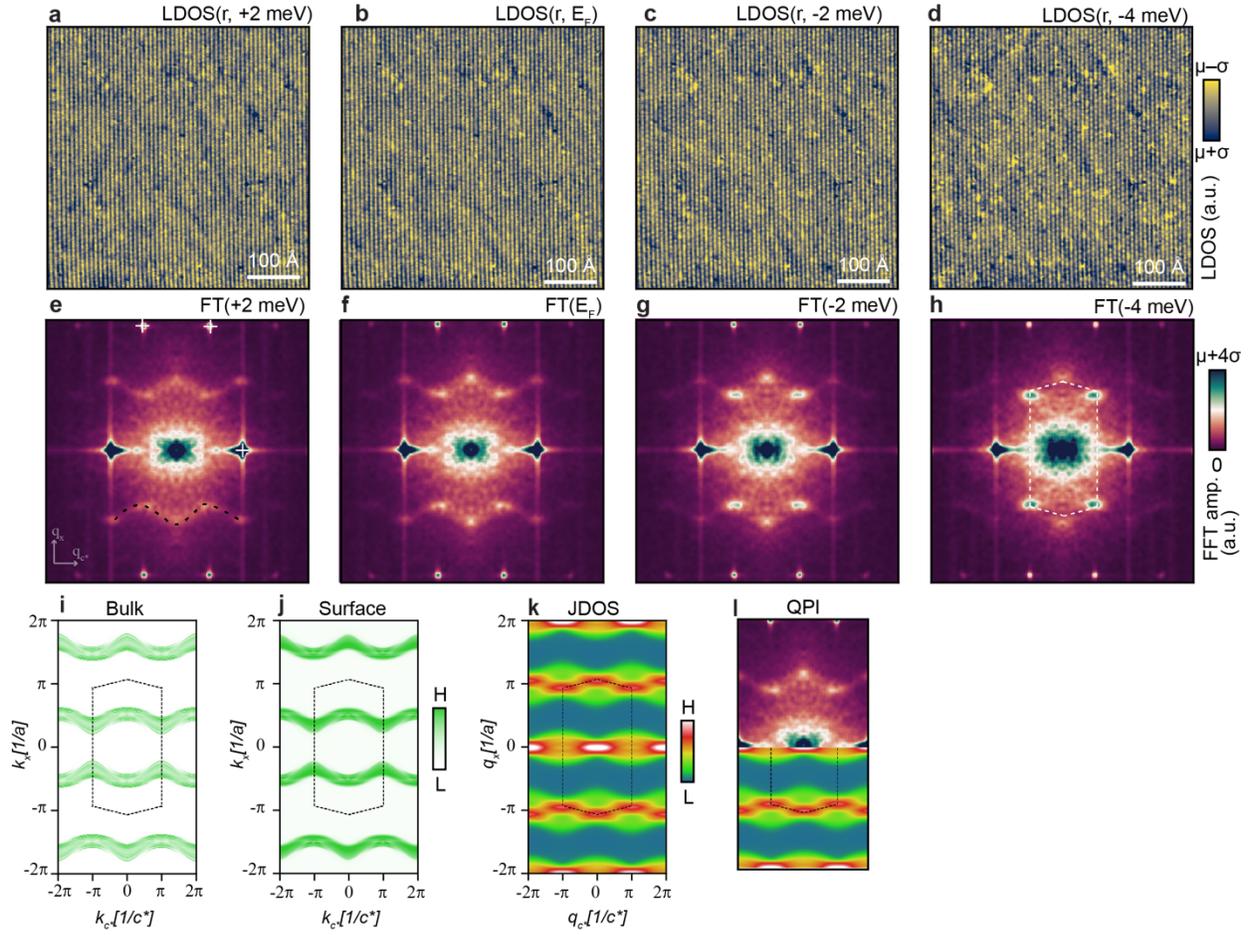

**Figure 3: Spectroscopic imaging and quasiparticle interference on the 0-11 surface of UTe2**

**a-d**, LDOS maps measured on the (0-11) surface at a few energies (2 meV, 0 meV, - 2 meV, -4 meV) obtained on a 50 nm by 50 nm area (cropped topography shown in Fig.1c) ($V_{Bias}$ = 50 mV, I = 250 pA, T = 300 mK). **e-h**, Corresponding Fourier transforms of the LDOS maps shown in **a-d** showing the QPI patterns. The reciprocal lattice vector positions are shown in white crosses in **e**. The black dashed lines are a guide to the eye for the QPI signal. The full dataset is in the supplement. **i,j**, Bulk (**i**) and surface (**j**) spectral functions along the (0-11) direction calculated from tight-binding model. The bands have a dominant d character from uranium. The surface and the bulk spectral functions look qualitatively similar. Joint density of states (JDOS) at $E_F$ calculated from the tight-binding Fermi surface. The complete datasets can be found in the supplement. **l**, Juxtaposition of the experimental QPI at $E_F$ and the predicted JDOS (**k**). Panel **h-l** show the surface Brillouin zone in dashed lines. The Fourier transforms have been mirror symmetrized and applied with a gaussian smoothening to obtain high signal to noise ratio.

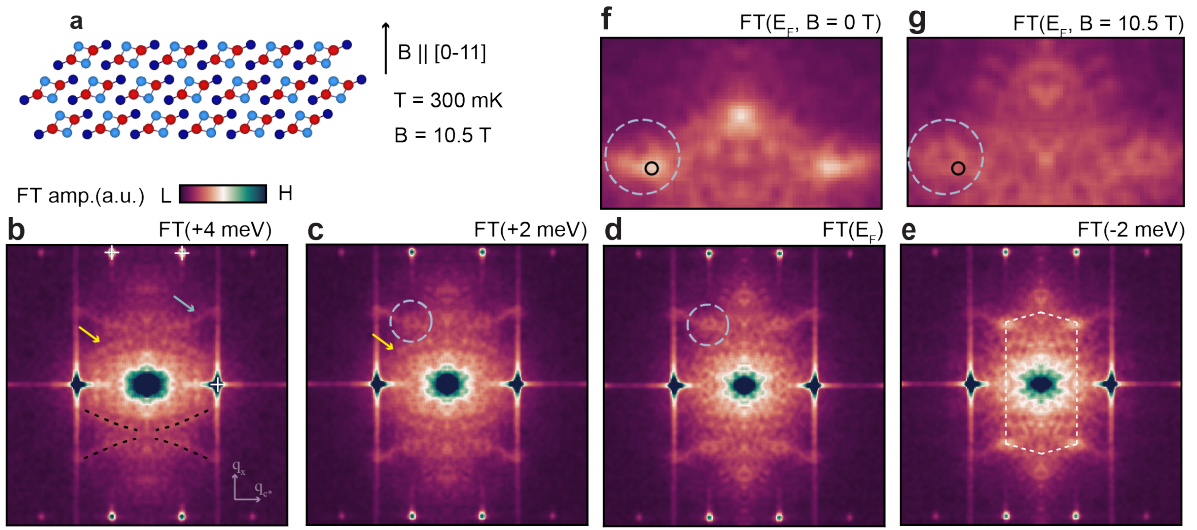

**Figure 4: QPI in high magnetic field near $E_F$ after suppressing the CDW**
**a,** Schematic of the (0-11) surface with an out of plane magnetic field. The applied field is parallel to the (0-11) direction. **b-e**, Fourier transforms of the LDOS maps at B = 10.5 T revealing the QPI patterns on suppression of the charge density wave at energies close to $E_F$. The LDOS maps for both these datasets were obtained on the same 50 nm by 50 nm area under the same conditions as Fig.3 ($V_{Bias}$ = 50 mV, I = 250 pA, T = 300 mK, B = 10.5 T). Under an applied magnetic field, we see circular pockets close to the Fermi energy at the corners of the Brillouin zone as shown by the dashed circles in panels **c** and **d**. **f,g**, Zoom-in of the FT of LDOS slices at $E_F$ comparing B= 0 T and B = 10.5 T datasets. We see the small circular pockets shown by the dashed circles, with and without the CDW peaks respectively.

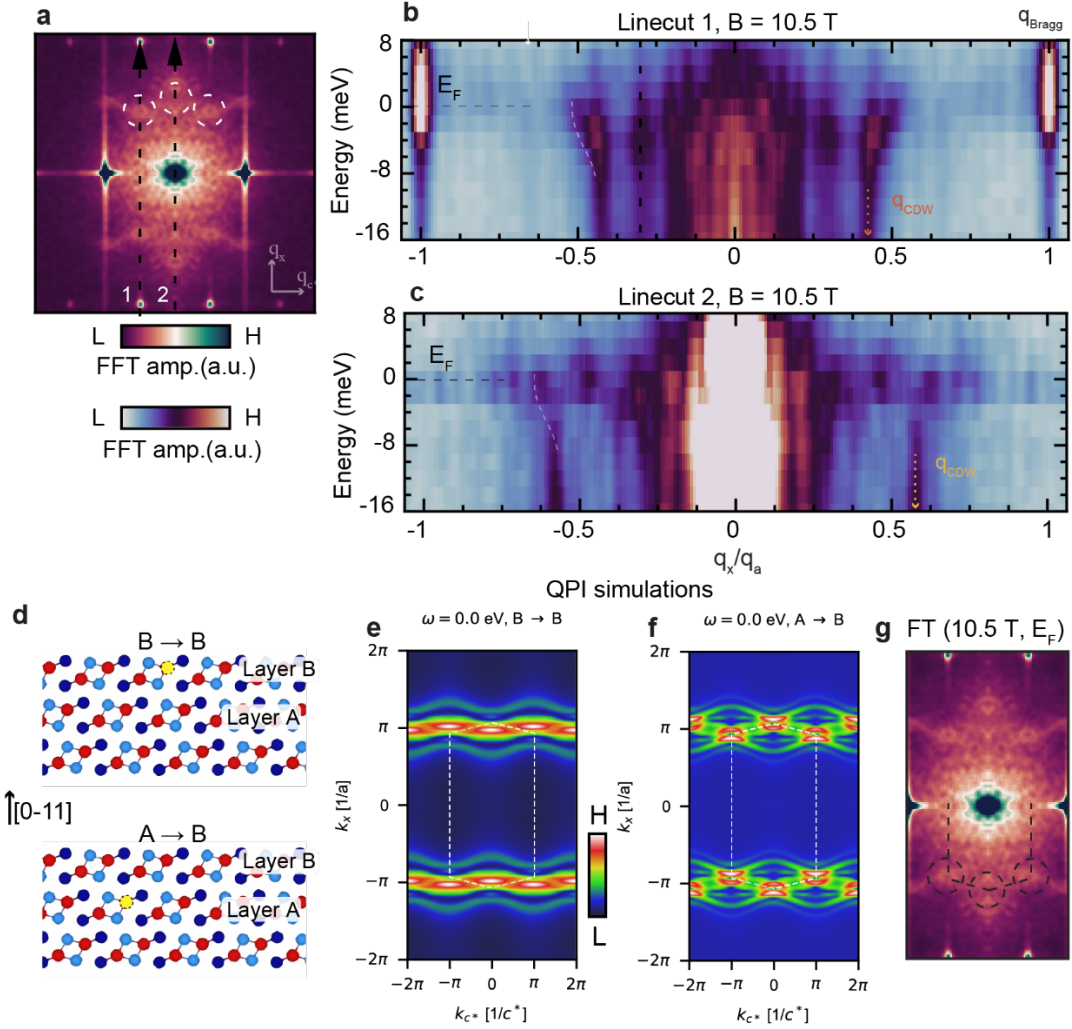

**Figure 5: Momentum-space cuts along different directions as a function of energy showing dispersive features**

**a,** FT of LDOS maps at $E_F$ B = 10.5 T elucidating the location of the linecuts (dashed black lines) labelled 1 and 2 shown in **b** and **c**. The circular Fermi pockets have been indicated by dashed white lines. **b-c,** Linecut 1 along $q_x$ at $q_{c^*} = -\frac{\pi}{c^*}$, at B = 10.5 T. We see a dispersing q-vector (white dashed curve in **b** and **c** are guide to the eye) close to $q_x = -0.43 \frac{2\pi}{c^*}$, which is the location of the CDW peak. **c,** Linecut 2 along $q_x$ at $q_{c^*} = 0$ at B = 10.5 T. The dashed black horizontal line indicates the location of the Fermi energy. **d**, Schematic showing the (0-11) direction with Layer B is the topmost layer and Layer A is the second layer. The top (bottom) panel denotes the scenario when the scattering center is on Layer B (Layer A). The position of a scattering defect at uranium site is indicated by a yellow dot. **e (f)**, QPI simulation of the (0-11) surface on Layer B when the scattering impurity is on Layer B (Layer A). **f** shows scattering intensity near the $(\pm\pi, \pm 1.1\pi)$ similar to what is seen in **a** and **g**. The dashed hexagon in **e** and **f** denote the SBZs. **g**, Comparison of experimental QPI at $E_F$ in the same q-space as **e-f** with a partial overlay of the SBZ and the circular pockets in black dashed lines.

**Acknowledgements**
The authors thank Daniel Agterberg for useful discussions. STM work at the University of Illinois, Urbana-Champaign was supported by the U.S. Department of Energy (DOE), Office of Science, Office of Basic Energy Sciences (BES), Materials Sciences and Engineering Division under Award No. DE-SC0022101. V.M. acknowledges support from the Gordon and Betty More Foundation's EPiQS Initiative through grant GBMF4860. H.C. acknowledges support from the Novo Nordisk Foundation grant NNF20OC0060019. A.K. acknowledges support by the Danish National Committee for Research Infrastructure (NUFI) through the ESS-Lighthouse Q-MAT. Research at the University of Maryland was supported by the Department of Energy Award No. DE-SC-0019154 (low-temperature sample characterization), the Gordon and Betty Moore Foundation's EPiQS Initiative through Grant No. GBMF9071 (materials synthesis), the National Science Foundation under the Division of Materials Research grant DMR-2105191, NIST, and the Maryland Quantum Materials Center.

**Author contributions**
A.A. and V.M. conceived the experiments. The single crystals were provided by S.R., S.R.S., J.P. and N.P.B. A.A. obtained the STM data. A.A. and V.M. performed the analysis and B.M.A., A.K and H.C performed the QPI simulations. A.A., and V.M. wrote the paper with input from all authors.

**Competing Interests**
The authors declare no competing interests.